\documentclass[conference,compsoc]{IEEEtran}
\IEEEoverridecommandlockouts
\usepackage{adjustbox}
\usepackage{algorithm}
\usepackage{algorithmic}
\usepackage{amsmath,amssymb,amsfonts}
\usepackage{array}
\usepackage{balance}
\usepackage{booktabs}
\usepackage[font=normalsize,skip=3pt]{caption}
\usepackage{cite}
\usepackage{colortbl}
\usepackage{dblfloatfix}
\usepackage{etoolbox}
\usepackage{filecontents}
\usepackage{graphicx}
\usepackage{hyphenat}
\usepackage{inputenc}
\usepackage{listings}
\usepackage{longtable}
\usepackage{lipsum}
\usepackage{makecell}
\usepackage{mdframed}
\usepackage{multirow}
\usepackage{parcolumns}
\usepackage{placeins}
\usepackage{subcaption}
\usepackage{tabularx}
\usepackage{textcomp}
\usepackage{tcolorbox}
\usepackage{tikz}
\usepackage{todonotes}
\usepackage{url}
\usepackage{xcolor}
\usepackage{float}
\usepackage{pdflscape}
\usepackage{seqsplit}
\usepackage{tabularray}
\usepackage{ragged2e}
\usepackage{array}

\definecolor{rank1}{RGB}{212, 237, 218}  % Light green - 1st place
\definecolor{rank2}{RGB}{179, 229, 252}  % Light blue - 2nd place  
\definecolor{rank3}{RGB}{255, 243, 205}  % Light yellow - 3rd place
\definecolor{rank4}{RGB}{248, 215, 218}  % Light red - 4th place

\usepackage{xcolor} % you likely already have this

\usepackage{listings}
\usepackage{xcolor}
\usepackage[hidelinks]{hyperref}

\lstdefinestyle{arrowstyle}{
  basicstyle=\footnotesize\ttfamily,      % smaller typewriter font
  breaklines=true,   % allow line wrapping
  breakanywhere=true, 
  breakautoindent=false
  numbers=left,
  keepspaces=true,
  numberstyle=\tiny,               % tiny font for line numbers
  frame=single,                    % draw a single-line border around code
  columns=fullflexible,            % better spacing for non-monospace fonts
  literate=
    {←}{{\ensuremath{\leftarrow}}}1
    {∅}{{\ensuremath{\varnothing}}}1
    {→}{{\ensuremath{\rightarrow}}}1
}
\def\BibTeX{{\rm B\kern-.05em{\sc i\kern-.025em b}\kern-.08em
    T\kern-.1667em\lower.7ex\hbox{E}\kern-.125emX}}
\begin{document}
\title{AndroByte: LLM-Driven Privacy Analysis through Bytecode Summarization and Dynamic Dataflow Call Graph Generation}

% \and
% \and

% }
% \author{
% \IEEEauthorblockN{Mst Eshita Khatun\IEEEauthorrefmark{1},
% Lamine Noureddine\IEEEauthorrefmark{1},
% Zhiyong Sui\IEEEauthorrefmark{1}, 
% and Aisha Ali-Gombe\IEEEauthorrefmark{1}}
% \IEEEauthorblockA{\IEEEauthorrefmark{1}Department of Computer Science and Engineering\\
% Louisiana State University, Baton Rouge, LA, USA \\
% mkhatu3@lsu.edu, lnoureddine@lsu.edu, zsui1@lsu.edu, aaligombe@lsu.edu}
% }
\author{
\IEEEauthorblockN{Mst Eshita Khatun*, 
Lamine Noureddine,
Zhiyong Sui, 
and Aisha Ali-Gombe}
\IEEEauthorblockA{Department of Computer Science and Engineering\\
Louisiana State University, Baton Rouge, LA, USA \\
mkhatu3@lsu.edu, lnoureddine@lsu.edu, zsui1@lsu.edu, aaligombe@lsu.edu}
\thanks{*Corresponding author: Mst Eshita Khatun }
}

\maketitle

\begin{abstract}
With the exponential growth in mobile applications, protecting user privacy has become even more crucial. Android applications are often known for collecting, storing, and sharing sensitive user information such as contacts, location, camera, and microphone data—often without the user’s clear consent or awareness—raising significant privacy risks and exposure.
%In privacy analysis, dataflow analysis is particularly valuable for identifying data usage and potential leaks.
%Traditionally, this type of analysis has relied on formal methods, heuristics, and rule-based reasoning. However, these techniques are often complex to implement and prone to errors, such as taint explosion for large programs. Moreover, most existing Android dataflow analysis methods depend heavily on predefined knowledge of sinks, limiting their flexibility and scalability. To address the limitations of these existing techniques, we propose AndroByte, an AI-driven privacy analysis tool that uses LLM for bytecode summarization to dynamically generate accurate data-flow call graphs from static code. AndroByte achieves a significant efficacy of F$\beta$-Score of 89\% in generating a dynamic dataflow call graph on the fly. Compared to traditional tools like FlowDroid and Amandroid, AndroByte demonstrates improved effectiveness in leak detection without dependency on predefined propagation rules or sink lists. Moreover, AndroByte's iterative approach to bytecode summarization provides comprehensive and explainable insights into dataflow and leak identification, achieving robust evaluation scores using the G-Eval framework. Overall, AndroByte's flexible design and effective integration of LLM reasoning significantly enhance its capability to provide scalable, accurate, and explainable privacy analysis for Android applications.
In the context of privacy assessment, dataflow analysis is particularly valuable for identifying data usage and potential leaks. Traditionally, this type of analysis has relied on formal methods, heuristics, and rule-based matching. However, these techniques are often complex to implement and prone to errors, such as taint explosion for large programs. Moreover, most existing Android dataflow analysis methods depend heavily on predefined list of sinks, limiting their flexibility and scalability. To address the limitations of these existing techniques, we propose AndroByte, an AI-driven privacy analysis tool that leverages the reasoning of a large language model (LLM) on bytecode summarization to dynamically generate accurate and explainable dataflow call graphs from static code analysis. AndroByte achieves a significant F$\beta$-Score of 89\% in generating dynamic dataflow call graphs on the fly, outperforming the effectiveness of traditional tools like FlowDroid and Amandroid in leak detection without relying on predefined propagation rules or sink lists. Moreover, AndroByte’s iterative bytecode summarization provides comprehensive and explainable insights into dataflow and leak detection, achieving high, quantifiable scores based on the G-Eval metric.
\end{abstract}

\begin{IEEEkeywords}
Privacy, Android, LLM, Dataflow, Call Graph, Bytecode Summarization
\end{IEEEkeywords}

\section{Introduction}\label{into}
In this digital age, mobile applications play a pivotal role, serving as indispensable tools that streamline our daily activities. According to Ericsson’s November 2022 Mobility Report, \cite{ericsson2022mobility}, there are more than 6.26 billion smartphone subscriptions worldwide, underscoring the central role that mobile technology plays in everyday life. Mobile applications often need access to user data for seamless operation, personalized user experiences, and enhanced functionality. For instance, apps such as navigation tools require location data to provide accurate directions, while streaming platforms analyze user preferences to suggest tailored content. Additionally, data collection supports app developers in optimizing performance through analytics, identifying bugs, and implementing user-driven updates. However, while these practices are essential for functionality and monetization, they highlight the need for transparency and robust privacy safeguards to protect users' sensitive information from misuse or unauthorized access. %Although most mobile operating platforms adopt user permissions to limit the use of sensitive personal information \cite{yang2022permdroid}, a significant number of mobile apps, particularly Android apps, tend to be overprivileged—requesting more data than necessary \cite{enck2014taintdroid,reardon201950}.

Regulations like the California Consumer Privacy Act (CCPA) as well as the General Data Protection Regulation (GDPR) in the European Union aim to safeguard user data. Furthermore, most mobile operating platforms adopt user permissions to limit the use of sensitive personal information \cite{yang2022permdroid}. Despite these multi-faced measures, a significant number of mobile applications, particularly Android applications, tend to be overprivileged, requesting more data than necessary \cite{enck2014taintdroid,reardon201950, luo2021automatic,wang2022runtime,bello2024user,bello2025the}. %Additionally, several other studies showed how Android applications' data collection practices can cause privacy leaks \cite{yang2013appintent,lo2014leakage,achara2014detecting,schindler2022privacy}.
In particular, several studies have shown that Android applications often engage in data collection practices that can lead to privacy leaks at the code-level behavior that are not disclosed by app developers \cite{yang2013appintent,lo2014leakage,achara2014detecting, hashmi2021longitudinal, tan2023ptpdroid,xiang2023policychecker, schindler2022privacy,morales2024large,wang2024you}. Therefore, it is essential to develop more robust code analysis tools capable of uncovering privacy risks that go beyond surface-level transparency measures.

To mitigate these code privacy-related issues, researchers in this domain have proposed numerous static and dynamic analysis techniques for dataflow analysis focusing on data leak detection. Taint analysis is a special type of dataflow analysis that follows the propagation of the target date from source to sink. Most dataflow analyses on Android are built on taint propagation leveraging either formal methods, heuristics, or strict rule-based approach \cite{arzt2014flowdroid,li2015iccta,wei2018amandroid}. However, with these traditional approaches, overly conservative taint propagation rules may mark many safe data flows as tainted, leading to high false positives and/or taint exposition, while strict predefined rules or heuristics may overlook certain data flows, resulting in missed detection. Additionally, formal method-based approaches are difficult to implement correctly, maintain, and adapt to evolving programming practices, such as new APIs or SDK versioning, while rule-based systems require significant manual effort to define and update rules for new libraries, APIs, or frameworks.

Thus, in this research, we proposed a novel technique called \texttt{\bf AndroByte} that addresses the critical challenge of dataflow analysis on Android. AndroByte's methodology leverages LLM's reasoning capability for method and dataflow analysis to identify privacy leaks in Android applications. At its core, AndroByte features a method analysis component based on bytecode summarization, which utilizes an integrated LLM to analyze the application's Smali code. This bytecode is parsed and summarized to generate natural language descriptions of the method behavior, focusing on the dataflow of the target data source and caller-callee relationships within the context of the given data source. Unlike traditional static analysis tools that rely on predefined rules or lists of sensitive data sinks, AndroByte dynamically identifies sinks through AI reasoning, enabling more adaptive and robust analysis. To determine the propagation path, AndroByte leverages its dynamic dataflow-aware call graph generation module, which constructs dataflow call graphs by recursively exploring the next method list based on the caller-callee relationship starting from the target data's API callsite. After the analysis, AndroByte provides a comprehensive summary of dataflows, the sensitive data propagation path, and the specifics of privacy leaks. The combination of LLM-driven reasoning, iterative graph generation, and method-level bytecode analysis allows AndroByte to address the limitations of traditional methods, offering a novel, automated, effective, and adaptive Android privacy analysis technique. Our evaluation of AndroByte in 300 real-world apps shows that our approach can generate accurate dataflow call graphs with 89\% F$\beta$-Score accuracy and provide comprehensive summaries that show propagation path, description, and sink information. Additional evaluation on DroidBench and UBCBench apps shows that AndroByte can automatically detect sinks without a predefined list data leak detection accuracy of F-Score 83\%. Additionally, the result of testing AndroByte on LLMs indicates that the proposed approach is model-agnostic, highlighting its robustness, portability, and adaptability across different LLMs.
% with G-Eval score.
\noindent In summary, the research makes the following key contributions: 
\begin{itemize} 
 \item Dynamic Dataflow-aware Call Graph Generation: AndroByte introduces an effective and automated mechanism for dynamically generating dataflow call graphs directly from bytecode summarization, leveraging AI reasoning to trace dataflows and caller-callee relationships iteratively.
 \item Robust and Adaptable Data Leak Detection: By eliminating the dependency on predefined rules and sink lists, AndroByte enables robust detection of data leaks through method analysis and dynamic graph generation, making it more adaptable than traditional approaches.
 \item Comprehensive Bytecode Summarization: AndroByte provides detailed natural language summaries for each method, highlighting sensitive data propagation paths and potential sinks.
 \item To support reproducibility and future benchmarking, we released AndroByte’s source code, complete dataflow call graph traces, summarization outputs, and all supplementary materials \cite{AndroByte_Artifacts_2025}. %\footnote{ The GitHub repository link is anonymized per ACSAC 2025 submission criteria and will be made public upon acceptance.}
\end{itemize}

\noindent \textbf{Paper outline.} The rest of the paper is outlined as follows: Section \ref{background} provides the background \& related work on privacy analysis with a specific emphasis on dataflow analysis; Section \ref{meth} offers a detailed description of AndroByte design; Sections \ref{implementation}, \ref{eval}, and \ref{discussion} present the specific of AndroByte implementation, testing and evaluation, and discussion of the results and Section \ref{conclusion} concludes the paper.

\section{Background \& Related Work}\label{background}

\begin{figure*}[!htbp]
\centering
\includegraphics[width=\textwidth,  height=0.23\textheight]{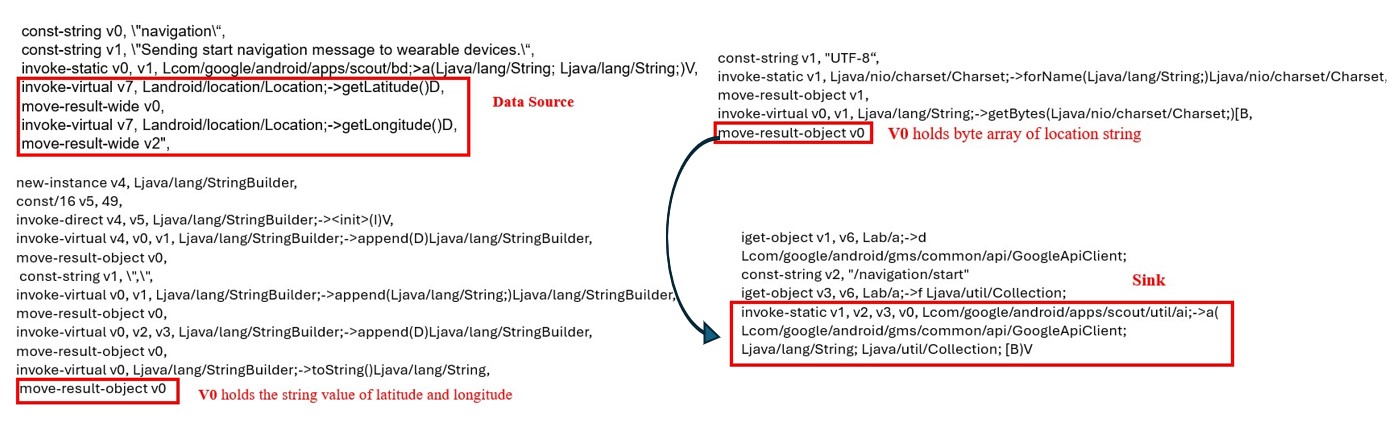}
\caption{Motivation Example}
\label{fig:motivation}
\end{figure*}

%\subsection{Background}
From an analysis perspective, the most reliable way to validate privacy exposure is through code-level privacy analysis, which examines data access, propagation, and potential leaks. While this analysis can be computationally expensive at runtime, it is more manageable when conducted statically. This process, commonly referred to as dataflow analysis, is more feasible in a static context. For instance, in Figure \ref{fig:motivation}, the bytecode instructions show that the method calls two sensitive APIs, \texttt{get latitude()} and \texttt{get longitude()} on a \texttt{android/location/Location} object. These calls retrieve the device’s current latitude and longitude, which are considered user sensitive data. The values are then appended to a comma-separated string (e.g., 'latitude, longitude') and stored in the register \texttt{v0}. Next, \texttt{v0} is converted into a UTF-8 byte array of the location coordinates. Finally, the method uses call \texttt{v1} is a GoogleApiClient, \texttt{v2} is the string "/navigation/start" \texttt{v3} is connected wearable devices, \texttt{v0} is the byte array containing location (latitude, longitude) to send this byte array outside of the app, specifically to a connected wearable device. In dataflow analysis, this call constitutes a sink because it is the point where the sensitive location data is sent outside the app’s immediate boundary.

%Traditionally, static dataflow analysis methodologies leverage taint propagation using formal methods, heuristics, and well-defined rules, which is very complex, particularly when applied to bytecode or assembly instructions. One well-known static taint analysis technique on Android built based on this traditional approach is FlowDroid\cite{arzt2014flowdroid}, is designed to require predefined data sources and sinks and then performs a context-, flow-, field-, and object-sensitive analysis of the app's bytecode to track the propagation of sensitive data. While many research papers have been developed to show FlowDroid's precision and effectiveness in analyzing complex dataflows, its methodology heavily relies on the precise definition of sink and rigid heuristics for taint propagation. In contrast, we proposed a novel architecture, \emph{AndroByte} that leverages program context reasoning using AI for the dataflow analysis and identifies the propagation path between source and sink without the complexities of traditional methods. Our approach is built on bytecode summarization and dynamic dataflow graph call generation using LLM capability.

%\subsection{Related work}
\subsection{Traditional Taint Analysis}
%Taint analysis is a specific technique within the broader framework of dataflow analysis, focusing on tracking the flow of sensitive (referred to as "tainted data") through a program. It identifies whether this data propagates to "sinks" where it could cause security or privacy issues. On Android, most dataflow analyses are built on taint propagation leveraging either formal methods, heuristics, or strict rule-based approach\cite{yang2012leakminer,octeau2013effective,arzt2014flowdroid,klieber2014android,bhosale2014precise,li2014automatically,li2015iccta,gordon2015information,wei2018amandroid,wei2018amandroid,bosu2017collusive}. For instance, FlowDroid, which is one of the most influential Android taint analysis system\cite{arzt2014flowdroid} conducts a flow-insensitive points-to analysis and leverages context-, flow-, field-, and object-sensitive analysis for taint propagation. EPICC\cite{octeau2013effective}, IccTA\cite{li2015iccta}, and Amandroid\cite{wei2018amandroid} extended FlowDroid capability to handle Android Intent objects. While these techniques have been shown to be effective, they heavily rely on predefined sinks and the complexities of predefined propagation rules. In contrast, our proposed AndroByte leverages Bytecode summarization via AI reasoning to automate the detection of sinks and the propagation of target data without the need for strict predefined rules.
Traditionally, static dataflow analysis methodologies leverage taint propagation to track the flow of sensitive data (referred to as "tainted data") through a program. These methods identify whether such data reaches "sinks" where it may cause security or privacy issues, using formal methods, heuristics, and well-defined rules \cite{yang2012leakminer,octeau2013effective,arzt2014flowdroid,klieber2014android,bhosale2014precise,li2014automatically,li2015iccta,gordon2015information,wei2018amandroid,bosu2017collusive, ali2016aspectdroid,ali2018toward, ali2017malware, ali2015opseq}. This process is particularly complex when applied to bytecode or assembly instructions. One of the most influential and widely used static taint analysis systems for Android, based on this traditional approach, is FlowDroid \cite{arzt2014flowdroid}. It requires predefined data sources and sinks, performs flow-insensitive points-to analysis, and incorporates context-, flow-, field-, and object-sensitive analysis for taint propagation. Tools like EPICC \cite{octeau2013effective}, IccTA \cite{li2015iccta}, and Amandroid \cite{wei2018amandroid} have extended FlowDroid's capabilities to support Android Intent objects.
While many studies have demonstrated FlowDroid’s precision and effectiveness in analyzing complex dataflows, its methodology heavily depends on rigid heuristics and precisely defined sinks for taint propagation. In contrast, we propose a novel architecture, AndroByte, which leverages AI-based program context reasoning for dataflow analysis. Our approach identifies propagation paths between sources and sinks without relying on the complexities of traditional static techniques. Instead, it builds on bytecode summarization and dynamic dataflow call graph generation using LLM capabilities.

\subsection{Bytecode summarization}
The advancements in language models have significantly enhanced automatic code summarization techniques to generate natural language descriptions that succinctly explain the code's purpose and functionality\cite{walton2024exploring}. Xiang et al. \cite{xiang2024automating} developed SmartBT, a summarization tool for Ethereum bytecode utilizing control flow graphs (CFGs) and information retrieval techniques. Similarly, BCGen \cite{huang2023bcgen} and further studies by Huang et al. \cite{huang2025towards} leveraged CFGs with semantic insights to effectively summarize Java bytecode. Cobra \cite{li2024cobra} applies bytecode summarization for enhanced vulnerability detection in smart contracts, while StoneDetector \cite{schafer2023finding} detects code clones by examining bytecode patterns. Fevid et al. \cite{fevid2024zero} introduced a Random Forest approach analyzing opcode sequences to detect ransomware threats at the assembly level. LTAChecker \cite{liu2024ltachecker} similarly employed opcode sequences for Android application behavior analysis. The key distinction between these related works and our proposed AndroByte  focuses on privacy analysis and D2CFG generation using AI-driven reasoning. To the best of our knowledge, our work is the first to propose leveraging LLM reasoning capability for method and dataflow analysis. 

\subsection{Privacy Analysis Using AI}
Recent studies have shown that integrating a machine learning-based approach with traditional static analysis methods can provide effective approaches to privacy analysis. For instance, Jain et al.\cite{jain2022pact} employed RNN models on Abstract Syntax Tree (AST) paths to detect privacy behaviors in Android code, while SAMLDroid \cite{hu2021samldroid} integrated RF with taint analysis to identify location privacy leakage. Feichtner et al. \cite{feichtner2020understanding} utilized CNNs and word embeddings to assess alignment between app descriptions and permissions, and Rahman et al. \cite{rahman2022permpress} leveraged FastText and BERT to identify permission-related content in policies. Similarly, Ma et al. \cite{ma2020deep} developed SideNet, a deep-learning framework detecting sensitive app activities through Encoder and ResNet-based models. Liu et al. \cite{liu2021nleu} introduced NLEU, enhancing taint analysis with NLP-derived semantic insights, and Fu et al. \cite{fu2017leaksemantic} employed a Decision Tree combined with a bag-of-words model to classify network transmissions as legitimate or illegitimate. Further, Morales et al. \cite{morales2024large} used LLMs and semantic similarity to detect inconsistencies between privacy policies and actual code practices, whereas Privacify \cite{woodring2024enhancing} applied LLM-based summarization for privacy policy comprehension, and PrivacyAsst \cite{zhang2024privacyasst} applied LLM-driven encryption and attribute shuffling to secure sensitive user data. In contrast, our approach leverages LLM reasoning to perform method-level bytecode and dataflow analysis specifically aimed at detecting data leaks in Android applications. 
.

\section {Design}\label{meth} %2 pages
\begin{figure*}[!htbp]
\centering
\includegraphics[width=0.7\textwidth, height=0.4\textheight]{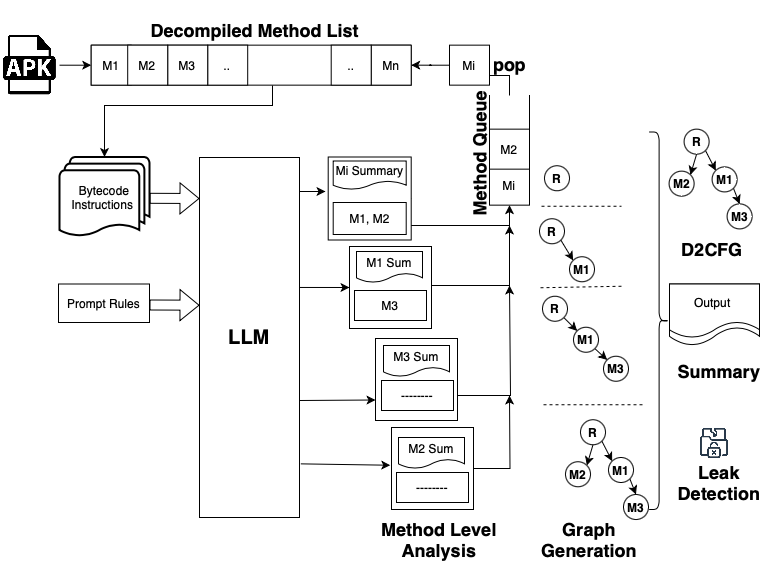}
\caption{AndroByte's Workflow}
\label{fig:workflow}
\end{figure*}

This section presents the design of AndroByte — a code-level privacy behavior analyzer for Android applications as illustrated in Figure \ref{fig:workflow}. AndroByte has three main components: 1) Method Analysis; 2) Prompt Engineering; and 3) Dynamic Dataflow Call Graph Generation (D2CFG).

%\begin{itemize}
%    \item Method Analysis
%    \item Prompt Engineering
%    \item Dynamic Dataflow Call Graph Generation (D2CFG)
%\end{itemize} 

% \begin{figure*}[!htbp]
% \centering
% % \includegraphics[width=\textwidth, height=0.45\textheight, keepaspectratio]{ACSAC 2025/Figures/AndroByte.png}
% \includegraphics[width=0.8\textwidth, height=0.4\textheight]{ACSAC 2025/Figures/AndroByte.png}
% \caption{AndroByte's Workflow}
% \label{fig:workflow}
% \end{figure*}

\subsection{Method Analysis}
The method analysis leverages bytecode summarization using LLMs to understand, analyze, and explain method-level behavior for a target application. Bytecode summarization was chosen over source code summarization primarily due to its platform-independent representation that reflects the executable logic of an Android application. This makes bytecode closer to the final executable form, providing a more reliable basis for understanding the app’s runtime behavior.
While obfuscation is common among Android developers to obscure the original source code, often affecting its readability and making decompilation less reliable, decompilation into bytecode largely retains the instructional structure of a program, thereby preserving key program context and dependency. Compared to source code decompilation, bytecode is more likely to maintain the actual instructions, method context, and dependencies needed for accurate analysis. Thus, in this component, we focus on bytecode summarization to analyze target functions in the context of sensitive data. The algorithm traverses the bytecode for a function to find the relationships between target data and other variables and trace the dataflow through the function's caller-callee relationship.\\
% While Android developers commonly apply obfuscation techniques to obscure original source code, making source code decompilation challenging and less reliable whereas bytecode remains structurally intact, preserving executable instructions and essential program logic. Consequently, bytecode analysis provides a more precise, robust, and reliable foundation for accurate data-flow analysis compared to source code analysis.
\noindent Let \( M \) represent a set of methods extracted from a decompiled Android application:
\[
M = \{m_1, m_2, m_3, \dots, m_n\}
\]
where \( m_i \) is the bytecode representation of the \(i\)-th method that calls a target data source API \(S_s \), and thus \( m_i \) is defined as:
\[
m_i = \{S_s, I, C, D\}
\] 
where:\\
- \( S_s\): The target data source.\\
- \( I \): Bytecode instructions.\\
- \( C \): Contextual information (e.g., class name, method signature, modifiers).\\
- \( D \): Dependency information (e.g., invoked methods or APIs).

\noindent The objective function of this method analysis component is to generate a structured and human-readable representation \( S(m_i) \) for each method \( m_i \), which includes:
\begin{enumerate}
    \item A natural language summary of the method in the context of \( S_s\).
    \item The list of all methods called (\( \texttt{nextMethods} \)) that have a caller-callee relationships with \( m_i \) in the context of \( S_s\) data flow.
    \item Identified sensitive data sinks in the context of \( S_s\).
\end{enumerate}

\noindent To achieve this objective, we describe the transformation process from \( m_i \) to \( S(m_i) \) as a composition of three sub-components:
\begin{itemize}
    \item Bytecode parsing module: Extracts the components of \( m_i \) from a decompiled app:
   \[
   P(m_i) \rightarrow \{S_s, I, C, D\}
   \]
   \item LLM summarization module:
   \begin{itemize}
       \item Automatically generates natural language ({$NL$}) summary using pre-trained model: 
           \[
            \mathcal{L}(P_{prompt}) \rightarrow S_{\text{NL}}(m_i)
            \]
      \item Identifies all methods called by \( m_i \) and represents them as \( \texttt{nextMethods} \), 
            \[
            \mathcal{C}(m_i) \rightarrow \texttt{nextMethods} = \{n_1, n_2, \dots, n_j\}
            \]
       \item Analyzes the flow of sensitive data within \( m_i \) from a target source (\( S_s \)) to sinks (\( S_k \)):
            \[
           \mathcal{D}(m_i) \rightarrow \{S_s, S_k\}
           \]
       \item Finally, the aggregation module combines all outputs into a structured representation \( S(m_i) \):
           \[
           S(m_i) = \{S_{\text{NL}}(m_i\rightarrow S_s,), \texttt{nextMethods}, S_k\}
           \]
    \end{itemize}
\end{itemize}

\subsection{Prompt Engineering}
The prompt engineering component is critical to ensuring the effectiveness of the bytecode summarization process in identifying data flow and determining the list of \texttt{nextMethods}. This component leverages structured techniques like scoping and few-shot prompting to optimize the prompts provided to the LLM, enhancing its ability to generate accurate, relevant, and efficient output.

\subsubsection{Scoping} This prompt engineering technique involves tailoring the prompt to include only the most relevant information in a target method's bytecode instructions. The components that are selectively prioritized to guide the LLM toward specific tasks are the target data origin (\( S_s \), the generic definition of bytecode instructions \( I \), the sink \( S_k \), the context of data flow propagation, storage and dependency \( C \), and \( D \). Thus, our Scoping function (\( \mathcal{S} \)), as shown below, extracts and prioritizes the most relevant information from \( P(m_i) \) to construct a targeted scope for the prompt:
   \[
   \mathcal{S}(P(m_i)) \rightarrow \{S_s', S_k', I', C', D'\}
   \]
   where \( S_s' \), \( S_k' \),\( I' \), \( C' \), and \( D' \) represent scoped subsets of \( S_s \), \( S_k \), \( I \), \( C \), and \( D \), respectively.
\subsubsection{Few-Shot Prompting}  This technique provides the LLM with a small set of examples within the prompt. These examples demonstrate the expected input-output relationship, improving the model’s contextual understanding. The examples are curated to reflect scenarios like identifying method signatures and parameters. Our Few-Shot prompt construction (\( \mathcal{F} \)) adds examples demonstrating the desired outcomes for summarization, next methods, and data flow analysis:
   \[
   \mathcal{F}(\{I', C', D'\}) \rightarrow E
   \]
   where \( E \) is the set of examples embedded in the prompt.

\noindent Hence, our prompt for the method analysis (\( \mathcal{P} \)) combines the scoped data and examples into a coherent, structured prompt:
   \[
   \mathcal{P}(\mathcal{S}, \mathcal{F}) \rightarrow P_{prompt}
   \]

\subsection{Dynamic Dataflow-aware Call Graph Generation - D2CFG}
This component builds a graph representation of the methods for the target Android application, where nodes represent methods and edges represent caller-callee relationships. This process leverages the results of method analysis and prompts engineering to recursively explore methods, adding nodes and edges to the graph until a termination condition is met - no new methods are found, or a viable sink is identified. The significance of this novel D2CFG is that it dynamically traces data sources and identifies sinks through the graph without the complexity of taint propagation rules. 

The objective function of the D2CFG is to construct a directed graph \( G \) where:\\

- \( N \): Nodes represent methods \( m_i \)

- \( E \): Edges represent caller-callee relationships between methods

\noindent The graph captures the flow of sensitive data and identifies potential sinks for \textbf{a given target data source}. Thus, the initial input to the graph is the root method, which is identified as \( S_s' \)'s callsite. Let's denote this root method  \( m_0 \).

To generate the D2CFG with \( m_0 \) as the root node, the following steps are taken:

1. Initialize an empty directed graph and add the starting method \( m_0 \) as the root node:
     \[
     G = (N, E), \quad N = \emptyset, \quad E = \emptyset
     \]
     \[
     N \leftarrow \{m_0\}
     \]

2. Create a queue \( Q \) and add \( m_0 \) to it:
     \[
     Q \leftarrow \{m_0\}
     \]

3. Recursive Exploration: While the queue \( Q \) is not empty:
\begin{itemize}
    \item Dequeue a method by removing the first method \( m_i \) from \( Q \).
    \item Summarize \( m_i \) using the method analysis and prompting modules to retrieve:
          \[
          \mathcal{L}(P_{prompt}) \rightarrow S_{\text{NL}}(m_i\rightarrow S_s), \texttt{nextMethods}, S_k
          \]
          \item Add \texttt{nextMethods} to Graph:
          \begin{itemize}
              \item For each \( n_j \in \texttt{nextMethods} \):
              \begin{itemize}
              \item Add \( n_j \) to \( N \) if it is not already in the graph:
                \[
                N \leftarrow N \cup \{n_j\}
                \]
               \item Add an edge from \( m_i \) to \( n_j \):
                \[
                E \leftarrow E \cup \{(m_i, n_j)\}
                \]
                 \end{itemize}
                \item Add \( n_j \) to \( Q \) for further exploration.
          \end{itemize}
           \item If a viable sink \( S_k \) is identified for the target data, then record the sink in the graph and, optionally, terminate further exploration of this path.              
\end{itemize}
4. This process ends when \( Q \) is empty, indicating no more methods to explore or when a sink is found for the target data.

Thus, with the method analysis extracting, summarizing, and identifying key relationships within individual methods, prompt engineering optimizes the method analysis's interactions with the LLM, ensuring precise and context-aware summarization. Finally, the dynamic data flow graph generation constructs a recursive and iterative representation of the program's behavior, connecting methods through caller-callee relationships and tracing sensitive data flow to identify potential sinks. \vspace{-0.2cm}

\section{Implementation}\label{implementation}
Based on the module design discussed above, we developed a Proof-Of-Concept (POC) for AndroByte. The POC has three modules for method analysis, prompt engineering, and dynamic dataflow graph generation. In this POC, we limit the source definition to the list of sensitive APIs that are commonly used in Android applications to collect privacy-related user information, as shown in Table \ref{tab:privacy-apis}.

The primary input for AndroByte's method analysis module is the decompiled Android code in \texttt{Smali} format. Smali is a human-readable intermediate representation of bytecode, closely resembling assembly instructions. To streamline this process, we integrated the APK decompiler tool, APKTool\cite{ibotpeaches2018apktool}, into AndroByte to convert the Dalvik bytecode from each APK into its Smali representation. At the start of the application analysis, all call sites for sensitive APIs listed in Table \ref{tab:privacy-apis} are identified and initialized as root nodes for the D2CFG. Starting from each root node, the method analysis module is recursively invoked to perform bytecode summarization, identify callee methods through the next methods list, and determine potential sinks. The prompts used in this process are scoped and fine-tuned to follow the context of the sensitive dataflow. An example of the method analysis prompt is shown in Appendix \ref{appA} as Listing \ref{lst:prompt}. The D2CFG generation module employs a queue-based recursive function to manage the analysis flow. Callee methods returned by the method analysis are added as child nodes to the graph, which is then traversed iteratively. This process continues until a viable sink is identified or all reachable methods have been analyzed. This recursive exploration ensures that all method's caller-callee relationships in the path of the target dataflow are accurately captured in the graph.

\begin{table}[t]
\caption{Examples of Android APIs Accessing Personal Data}
\renewcommand{\arraystretch}{1.2}
\setlength{\tabcolsep}{4pt} % Reduce padding between columns
% \scriptsize % Shrink font to help with fit
% \footnotesize
\begin{tabular}{|>{\raggedright\arraybackslash}p{6cm}|>{\raggedright\arraybackslash}p{2cm}|}
\hline
\textbf{API Name} & \textbf{Data Type} \\ \hline
\seqsplit{android/location/Location;getLatitude()} & Location \\ \hline
\seqsplit{android/location/Location;getLongitude()} & Location \\ \hline
\seqsplit{android/telephony/TelephonyManager;getLine1Number()} & Phone Number \\ \hline
\seqsplit{android/accounts/AccountManager;getAccounts()} & Email Address \\ \hline
\seqsplit{android/telephony/TelephonyManager;getDeviceId()} & Device Identifier \\ \hline
\seqsplit{android/telephony/TelephonyManager;getSimSerialNumber()} & SIM Serial \\ \hline
\seqsplit{android/net/wifi/WifiInfo;getMacAddress()} & MAC Address \\ \hline
\seqsplit{android/net/wifi/WifiInfo;getSSID()} & SSID \\ \hline
\seqsplit{android/net/wifi/WifiInfo;getBSSID()} & BSSID \\ \hline
\seqsplit{android/location/LocationManager;getLastKnownLocation()} & Location \\ \hline
\seqsplit{android/telephony/SmsMessage;getDisplayMessageBody()} & Message Content \\ \hline
\end{tabular}
\label{tab:privacy-apis}
\end{table}
% \begin{figure*}[!t]
% \centering
% %\includegraphics[width=\textwidth, height=0.23\textheight]{ACSAC 2025/Figures/AndroByteGraph.png}
%     \includegraphics[width=\textwidth, height=0.23\textheight]{ACSAC 2025/Figures/AndroByteGraph_betterView.png}
%  \caption{D2CFG generated by AndroByte for Meme Generator app }
% \label{fig:memeApp}
% \end{figure*}
\begin{figure*}[!t]
  \centering
  \includegraphics[width=\textwidth, height= 0.3\textheight]{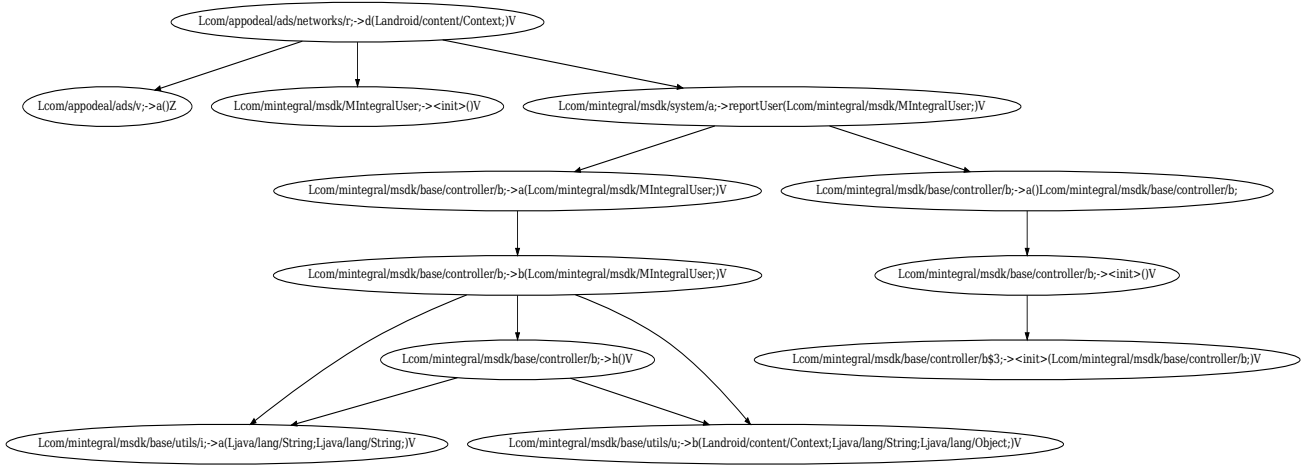}
  \caption{D2CFG generated by AndroByte for Meme Generator app}
  \label{fig:memeApp}
\end{figure*}

To illustrate the D2CFG graph generation, we provide an example in Figure \ref{fig:memeApp} for the "Meme Generator" app. At method analysis initialization, AndroByte found the callsite for the sensitive API \texttt{getLatitude()} as \lstinline{Lcom/appodeal/ads/networks/r;->d:(Landroid/content/Context;)V} and placed the method as root node in a empty graph. Following the method analysis for the root node, AndroByte found three nextMethods within the dataflow context of the location data, which are then added as nodes to the D2CFG with edges connecting these methods to the root node. Traversing these methods using the queue-based approach shows that two of the three nodes did not propagate the location data further, but the \lstinline{Lcom/mintegral/msdk/system/a;->reportUser:(Lcom/mintegral/msdk/MIntegralUser;)V} function passes the location data to the \lstinline{Lcom/mintegral/msdk/base/controller/b;->a:(Lcom/mintegral/msdk/MIntegralUser;)V} function. This function then calls the \lstinline{Lcom/mintegral/msdk/base/controller/b;->b(Lcom/mintegral/msdk/MIntegralUser;)V}, which processed the user object by serializing it to JSON and logs the location data via the \texttt{Log.d} utility in the \lstinline{Lcom/mintegral/msdk/base/utils/i;->a:(Ljava/lang/String; Ljava/lang/String;)V} directly and persist the JSON and associated (\texttt{expiretime}) using the method \lstinline{Lcom/mintegral/msdk/base/utils/u;->b}.
\section{Testing and Evaluation}\label{eval}
% Now that we have established the design of AndroByte and how it works using an example, in this section, we will evaluate the performance of our proposed approach based on the following research questions:
In this section, we evaluate the performance of our proposed approach based on the following research questions:
\begin{itemize}
% \item \textbf{RQ1 -} How accurate is AndroByte in identifying data leaks using different LLM configurations with D2CFGs, compared to traditional static analysis methods that rely on predefined sink lists?
\item \textbf{RQ1 -} How accurate is AndroByte in identifying data leaks with D2CFGs using different  LLM configurations, compared to traditional static analysis tools that rely on predefined sink lists?
\item \textbf{RQ2 -} How effective is AndroByte in generating automated D2CFG through the dynamic summarization of bytecode instructions?
\item \textbf{RQ3 -} How explainable is AndroByte’s method analysis in tracking sensitive dataflow and revealing potential data leaks in Android applications?
% \item \textbf{RQ4 -} How portable and adaptable is AndroByte's design across different LLMs?
\end{itemize}
 
\subsection{Experiment Setup}
\subsubsection{Benchmarks} \label{dataset}
To answer the above research questions, we selected two privacy benchmark suites that are commonly used in previous studies\cite{wei2018amandroid,grech2017p, mitra2019benchpress, chen2025pacdroid}, DroidBench \cite{arzt2014flowdroid} and the more recent UBCBench  \cite{zhang2021analyzing}, both of which are designed to evaluate privacy leak scenarios in Android applications. These benchmarks cover a wide range of leakage patterns, including aliasing, implicit flows, lifecycle handling, reflection, field and object sensitivity, and app-specific behaviors such as callbacks, array manipulations, and emulator detection mechanisms. Given that AndroByte is designed to detect privacy leaks through sensitive API calls, we applied a filtering strategy to include only the test cases relevant to its detection scope. Specifically, test apps in DroidBench that rely on non-API-based flows—such as hardcoded values, user inputs without sensitive API interactions, or leaks propagated through ICC were excluded from the evaluation. Similarly, although UBCBench contains 30 test cases, we retained only those that exhibit privacy behavior. As a result, from the original 118 apps in DroidBench and 30 in UBCBench, we selected 86 and 24 test cases, 110 in total for our experiments in RQ1 and RQ3. Additionally, we collected 350 real-time Android applications from the Google Play Store during July 2024, focusing on privacy-relevant categories such as social networking, health, finance, and other domains where sensitive user data are commonly accessed. As mentioned in the contribution, our overarching objective of this research is to build a system that automatically detects dataflows and potential data leaks using LLM's summarization capability. For experimental evaluation of RQ2 and RQ3, we focused on 11 representative sensitive APIs (as listed in Table~\ref{tab:privacy-apis}), and selected 300 out of 350 Apps that invoke at least one or more of these APIs. It is important to note that our approach is not inherently limited to this predefined set; the selected APIs serve as a practical subset for validating the system's effectiveness in real cases.

\subsubsection{LLM Model selection:} 
Recent advancement of LLMs have significantly expanded their functional scope, enabling enhanced performance in a wide range of engineering tasks. For the privacy analysis of mobile applications, we initially selected four LLMs from the open-source platform Ollama based on several criteria, including architectural diversity—to ensure balanced and unbiased evaluation—and model size, to balance computational efficiency without requiring large-scale infrastructure. The selected models range from 8 to 15 billion parameters. Table \ref{tab:llm_models} represents the overview of selected models.

\begin{table}[ht]
\centering
\caption{Overview of Selected Large Language Models}
\resizebox{\columnwidth}{!}{%
\begin{tabular}{lllll}
\hline
\textbf{Model Name} & \textbf{Architecture} & \textbf{Context Length} & \textbf{Quantization} & \textbf{Release Date} \\
\hline
qwen3:latest & Qwen3 & 40960 & Q4\_K\_M & April 2025 \\
llama3.1:latest  & llama & 131072 & Q4\_K\_M & July 2024 \\
deepseek-coder-v2:16b & deepseek2 & 163840 & Q4\_0 & June 2024 \\
gemma3:latest & gemma3 & 131072 & Q4\_K\_M & March 2025 \\
\hline
\end{tabular}
}
\label{tab:llm_models}
\end{table}
Additionally, all selected models support low-bit quantization techniques, which significantly reduce resource consumption and make it possible to run the models efficiently. Our selection emphasizes recently released models to ensure that the evaluation reflects the latest improvements in architecture, training strategies, and real-world applicability. To maintain consistency, all models were evaluated with a fixed context length of 40,000 tokens and a generation temperature of 0.2 to reduce variability.

% \noindent \textbf{Llama}: LLaMA \cite{touvron2023llama}is Meta’s open-source language model family, available in sizes up to 405B parameters. It excels in general reasoning, multilingual tasks, and long-context applications.\\
% \noindent\textbf{Gemma}: Gemma \cite{gemma2024open}, developed by Google, is a lightweight LLM series optimized for code generation, math, and efficient inference. The latest version scales up to 27B parameters.\\
% \noindent\textbf{Qwen}: Qwen \cite{qwen2023technical} is a multilingual model suite from Alibaba, designed for tasks such as translation, reasoning, and coding. Qwen3 features a broad lineup of both dense and Mixture-of-Experts models. \\
% \noindent\textbf{DeepSeek}:DeepSeek \cite{deepseek2024v3} is an open-source model family focused on high-accuracy code and math generation. Deepseek coder model is designed for code specific tasks.

%For our experiments, all models were sourced from publicly available checkpoints on Ollama, an open-source platform that facilitates local deployment, customization, and sharing of LLMs. 

% \vspace{-0.2cm}
% \subsubsection{Hardware Configuration} All experiments were conducted on a local server features an NVIDIA GeForce RTX 4090 GPU, an Intel Core i9-13900KF CPU, and 64GB of RAM, ensuring optimal performance for handling LLMs.
\subsubsection{Hardware Configuration} All experiments were run on a local server with an RTX 4090 GPU, Intel i9-13900KF CPU, and 64GB RAM.

\subsection{Evaluation Results}
\subsubsection{RQ1-  How accurate is AndroByte in identifying data leaks with D2CFGs using different  LLM configurations, compared to traditional static analysis tools that rely on predefined sink lists?}
In this evaluation, our primary goal is to examine the accuracy of AndroByte in identifying
sensitive dataflows and detecting data leaks using the D2CFG. 
%This metric is crucial for assessing the overall effectiveness of AndroByte design from its method analysis that conducts the summarization to the prompt engineering, which guides the model's summarization, and the D2CFG, which is used to detect the flow of sensitive data and identifies potential sinks. 
This metric is crucial for assessing the overall effectiveness of the AndroByte design. It covers the method analysis that performs code summarization, the prompt engineering that guides the model’s behavior, and the D2CFG construction. The D2CFG is used then to trace the flow of sensitive data and identify potential sinks.

\begin{table*}[ht]
\centering
\caption{Leak Detection Results for AndroByte with Different LLMs Compared to Amandroid and FlowDroid}
 % \scriptsize
\footnotesize

\setlength\tabcolsep{1.2pt} % reduces horizontal padding between columns
\renewcommand{\arraystretch}{1.2}
\begin{tabular}{|l|c|
ccc|ccc|ccc|ccc| % AndroByte LLMs
ccc|ccc|}        % Amandroid + FlowDroid
\hline
\multirow{3}{*}{\textbf{TestCases Suite}} & \multirow{3}{*}{\textbf{GT \#Leak}} 
& \multicolumn{12}{c|}{\textbf{AndroByte (LLM Variants)}} 
& \multicolumn{3}{c|}{\textbf{Amandroid }} 
& \multicolumn{3}{c|}{\textbf{FlowDroid}} \\
\cline{3-14}
& & \multicolumn{3}{c|}{\textbf{Gemma3}} 
  & \multicolumn{3}{c|}{\textbf{LLaMA3}} 
  & \multicolumn{3}{c|}{\textbf{Qwen3}} 
  & \multicolumn{3}{c|}{\textbf{DeepSeekCoder}} 
  & & & 
  & & & \\
\cline{3-20}
& & \textbf{TP} & \textbf{FN} & \textbf{FP}
  & \textbf{TP} & \textbf{FN} & \textbf{FP}
  & \textbf{TP} & \textbf{FN} & \textbf{FP}
  & \textbf{TP} & \textbf{FN} & \textbf{FP}
  & \textbf{TP} & \textbf{FN} & \textbf{FP}
  & \textbf{TP} & \textbf{FN} & \textbf{FP} \\
\hline
Aliasing & 0 &  0  & 0  & 1 & 0  & 0  & 1 & 0 & 0 & 1 & 0  &  0 & 1 &  0 & 0  & 1 & 0  &  0 & 1 \\
AndroidSpecific & 9 & 8 & 1 &  0 & 7 & 1 & 0 & 7 & 2 & 0 & 7 & 2 &  0 & 6 & 3 &  0 & 7 & 2 & 0  \\
ArrayAndLists & 3 & 3 &  0 & 2 & 3 & 0 & 4 & 2 & 0 & 1 & 3 & 0 & 1 & 3 & 2 & 2 & 3 & 0 & 4 \\
Callbacks & 17 & 9 & 8 & 1 & 9 & 7 & 1 & 8 & 8 & 2 & 9 & 8 & 3 & 11 & 8 & 2 & 16 & 1 & 2 \\
Emulator detection & 6 & 4 & 2 &  0 & 4 & 2 & 0  & 5 & 1 & 0 & 3 & 2 & 0  & 4 & 2 & 0 & 4 & 2 & 0 \\
FieldAndObjectSensitivity & 2 & 1 & 1 &  0 & 1 & 1 & 0  & 1 & 0 & 3 & 1 & 1 & 4 & 2 & 0 & 0 & 2 & 0 & 0 \\
General Java & 20 & 19 & 1 & 1 & 16 & 5 & 1 & 15 & 5 & 1 & 14 & 5 & 2 & 13 & 7 & 3 & 13 & 7 & 5 \\
ImplicitFlow & 2 & 1 & 1 & 0  & 1 & 1 & 0  & 1 & 1 & 0 & 1 & 1 & 0 & 0 & 2 & 0 & 0 & 2 & 0 \\
LifeCycle & 17 & 12 & 5 &  0 & 11 & 6 & 0 & 11 & 6 & 0 & 12 & 3 &  0 & 10 & 7 & 0 & 13 & 4 & 0 \\
Reflection & 4 & 3 & 1 &  0 & 3 & 1 & 0  & 3 & 1 & 0 & 3 & 1 &  0 & 3 & 1 & 0 & 2 & 2 & 0 \\
UBCB & 24 & 18 & 6 & 0  & 18 & 6 & 1 & 17 & 7 & 3 & 17 & 7 & 1 & 13 & 9 & 3 & 15 & 9 & 1 \\
\hline
\textbf{Total} & 104 & 78 & 26 & 5 & 73 & 30 & 8 & 70 & 31 & 11 & 70 & 30 & 12 & 65 & 41 & 11 & 75 & 29 & 13 \\
\hline
\textbf{Precision, P = TP/(TP+FP)} &  & 93.98\% &  &  & 90.12\% &  &  & 86.42\% &  &  & 85.37\% &  &  & 85.53\% &  &  & 85.23\% &  &  \\
\textbf{Recall, R= TP/(TP+FN)} &     & 75.00\% &  &  & 70.87\% &  &  & 69.31\% &  &  & 70.00\% &  &  & 61.32\% &  &  & 72.12\% &  &  \\
\textbf{F1-score = 2PR/(P+R)} &   & 83.42\% &  &  & 79.35\% &  &  & 76.92\% &  &  & 76.92\% &  &  & 71.43\% &  &  & 78.13\% &  &  \\
\hline
\end{tabular}
\label{tab:leak_detection_comparison}

%\vspace{-0.2em} % optional spacing before the note
\caption*{\footnotesize \textbf{Note:}  This benchmark includes multiple suites, each composed of test applications designed to evaluate privacy leak detection capabilities. GT \# Leak refers to ground truth nnumber of leak of the benchmarks. In this context, a true positive (TP) is a correctly detected privacy leak, a false positive (FP) is an incorrect leak report, and a false negative (FN) is a missed leak.}
\end{table*} 

We evaluated the performance of AndroByte integrated with four distinct LLMs to determine which configuration yields the most effective privacy leak detection. The objective is to benchmark and identify the most suitable LLM for integration within AndroByte’s summarization module analysis pipeline. To compare the performance of AndroBye against traditional tools, we also included two established static analysis tools Amandroid \cite{wei2018amandroid} and FlowDroid \cite{arzt2014flowdroid}. The evaluation was conducted using a comprehensive set of test cases from the benchmarks as described in Section \ref{dataset}.
Table \ref{tab:leak_detection_comparison} show the comparative analysis between AndroByte configured with different LLMs and the traditional static analysis tools, Amandroid and FlowDroid. It is important to note that while Amandroid and FlowDroid require a predefined sink list (comprising 82 sink methods), AndroByte automatically identifies sinks based on its method contextual analysis.

The evaluation results, presented in Table 3, show that AndroByte integrated with the Gemma3 model achieved the highest precision of 93.98\%, recall of 75\%, and F1-score of 83.42\%. On the other hand, LLaMA3 offered a balanced trade-off with a recall of 70.87\% and a reasonable precision of 90.12\%. Qwen3 and DeepSeekCoder exhibited slightly lower recall but maintained competitive F1 Scores of 76.92\% each, indicating their ability to generalize across categories with moderate false-positive rates. In contrast, the traditional tools Amandroid and FlowDroid demonstrated lower recall 61.32\% and 72.12\%, respectively —despite comparable precision. These results highlight AndroByte’s advantage in identifying privacy leaks more accurately when guided by LLM-generated summaries, as validated by manual inspection, despite not relying on complex predefined taint propagation rules. More so, AndroByte's non-dependency on a sink list gave it a significant advantage and greater flexibility over static tools like Amandroid and FlowDroid. It is worth noting that while traditional tools rely heavily on predefined source lists for both identifying and tracing sensitive data flows, AndroByte uses the source list solely for selecting root nodes. The LLM-driven analysis then operates independently of predefined source or sink lists, enabling a broader semantic understanding of privacy-relevant behavior.
%as confirmed by manual inspection,%

Manual analysis of the apps missed by AndroByte showed that our proposed LLM-driven algorithm misses leaks in apps with multiple leaks originating from a single source method, thereby classifying them as a single leak. As a result, it does not traverse the secondary path and, therefore, doesn't trace to the second sink as shown in the example app FactoryMethods1 (from DroidBench) summary provided in Appendix \ref{appD} listing \ref{lst:leak1}. 
Additionally, we found that AndroBytes didn't perform well with event-driven, complex Android lifecycle-dependent applications, such as the Button2, ActivityLifecycle2, BroadcastReceiverLifecycle2, and FragmentLifecycle2 apps in DroidBench. Nonetheless, for an analysis system based on AI reasoning without the complexity of taint propagation rules and predefined sink, AndroByte's results are promising.
We believe that AndroByte's algorithm can be significantly improved by fine-tuning the next method selection strategy and ultimately enhancing the D2CFG.

% In Summary, AndroByte achieves higher recall than traditional tools like Amandroid and FlowDroid in identifying data leaks while offering an automated mechanism for sink detection. Notably, it eliminates the reliance on predefined complex propagation rules and sink lists, making it a more adaptable solution.
% \begin{mdframed}[backgroundcolor=gray!10, linecolor=blue!75!black]
\begin{mdframed}[backgroundcolor=gray!10]
In Summary, AndroByte achieves higher recall than traditional tools like Amandroid and FlowDroid in identifying data leaks while offering an automated mechanism for sink detection. Notably, it eliminates the reliance on predefined complex propagation rules and sink lists, making it a more adaptable solution.
\end{mdframed}

\subsubsection{RQ2- How effective is AndroByte in generating automated D2CFG through the dynamic summarization of bytecode instructions?} \label{tab:rq2}
In this evaluation, our primary goal is to assess the effectiveness of AndroByte in generating dynamic data-flow-aware call graphs, starting from the call site of a single data source. The evaluation focuses on the system's summarization capability to detect caller-callee relationships, identify potential subsequent methods in the call graph based on the sensitive data flow, perform method analysis with recursive tracking, dynamically add nodes and edges, and utilize a prompting strategy that fine-tunes and guides the summarization process. Leveraging a widely-used static analysis tool Androguard's \cite{desnos2018androguard} call graph generation as the ground truth, we validated the graphs generated from the 300 real-world applications in our dataset based on the Table \ref{tab:privacy-apis} mentioned sensitive APIs. This resulted in 300 large graphs with sub-graphs indicating the flow for the 11 sensitive APIs for our AndroByte's. We generated Androguard call graphs starting from the sensitive API call sites as the roots. We used precision, recall, F1-Score, and F$\beta$-Score (weighted harmonic mean) as metrics to evaluate effectiveness. In this analysis, we compare the edges—represented as (caller, callee) pairs—in the AndroByte-generated graph \( G_{\text{AB}}\) with those in the Androguard-generated graph \ \( G_{\text{AG}} \). The parameters are described as follows:
% True Positive: Edges that appear in both \( G_{\text{AG}} \) and \( G_{\text{AB}} \), \(\textbf{G}_{\textbf{AG}} \cap \textbf{G}_{\textbf{AB}}\)
% False Positive: Edges that appear in \( G_{\text{AB}} \) but are not found in \( G_{\text{AG}} \), \(\textbf{G}_{\textbf{AB}} \setminus \textbf{G}_{\textbf{AG}} \)
% False Negative: Edges that are missed by \( G_{\text{AB}} \) but are found in \( G_{\text{AG}} \), \( \textbf{G}_{\textbf{AG}} \setminus \textbf{G}_{\textbf{AB}} \)
% True Positive (TP): Edges that appear in both $G_{\text{AG}}$ and $G_{\text{AB}}$, $G_{\text{AG}} \cap G_{\text{AB}}$.\\
% False Positive (FP): Edges present in $G_{\text{AB}}$ but absent in $G_{\text{AG}}$, $G_{\text{AB}} \setminus G_{\text{AG}}$.
% False Negative (FN): Edges present in $G_{\text{AG}}$ but missing in $G_{\text{AB}}$, $G_{\text{AG}} \setminus G_{\text{AB}}$.

\noindent True Positive (TP): Edges that appear in both \( G_{\text{AG}} \) and \( G_{\text{AB}} \). 
\[
\textbf{TP} = \textbf{G}_{\textbf{AG}} \cap \textbf{G}_{\textbf{AB}} 
\]
False Positive (FP): Edges that appear in \( G_{\text{AB}} \) but are not found in \( G_{\text{AG}} \). \vspace{-.2cm}
\[
\textbf{FP} = \textbf{G}_{\textbf{AB}} \setminus \textbf{G}_{\textbf{AG}} 
\]
False Negative (FN): Edges that are missed by \( G_{\text{AB}} \) but are found in \( G_{\text{AG}} \). 
\[
\textbf{FN} = \textbf{G}_{\textbf{AG}} \setminus \textbf{G}_{\textbf{AB}} 
\]

 % \vspace{-0.4em}

% \textbf{Precision:} The proportion of correctly predicted edges to the total predicted edges 
% \[
% \text{Precision} = \frac{\text{TP}}{\text{TP} + \text{FP}}
% \]
% \textbf{Recall:} The proportion of correctly predicted edges to the total actual edges 
% \[
% \text{Recall} = \frac{\text{TP}}{\text{TP} + \text{FN}}
% \]
% \textbf{F1-Score: } The harmonic mean of precision and recall
% \[
% \text{F1-Score} = 2 \cdot \frac{\text{Precision} \cdot \text{Recall}}{\text{Precision} + \text{Recall}}
% \]
% \textbf{F$\beta$-Score: } The weighted harmonic mean of precision and recall
% \[
% F_\beta = (1 + \beta^2) \cdot \frac{\text{Precision} \cdot \text{Recall}}{(\beta^2 \cdot \text{Precision}) + \text{Recall}}
% \]
\begin{table}[ht]
\centering
\caption{AndroByte Graph generation Performance}
\renewcommand{\arraystretch}{1.2}
\begin{tabular}{|l|c|}
\hline
\textbf{Metric} & \textbf{Value (\%)} \\ \hline
Precision       & 91.75               \\ \hline
Recall          & 79.67               \\ \hline
F1-Score        & 85.25               \\ \hline
F\textsubscript{$\beta$}-Score (\(\beta=0.5\)) & 89.04       \\ \hline
\end{tabular}
\label{tab:performance_metrics}
\vspace{0.5em}

% \parbox{0.95\linewidth}{
\footnotesize
\textbf{Note:} Precision is the ratio of correctly predicted edges to all predicted edges; recall is the ratio of correctly predicted edges to all actual edges. F1-score is the harmonic mean of precision and recall. The F\textsubscript{$\beta$}-Score is calculated as:
\(
F_{\beta} = (1+\beta^2) \cdot \frac{\text{Precision} \cdot \text{Recall}}{(\beta^2 \cdot \text{Precision}) + \text{Recall}}
\)
% }
\end{table}

% \vspace{-0.2cm}
Table \ref{tab:performance_metrics} shows that AndroByte achieved a precision of 91.75\%, which is the proportion of correctly predicted caller-callee edges to the total predicted edges. The high precision value further validates AndroByte's claim of correctly generating dynamic D2CFG on the fly using summarization capability. On the other hand, AndroByte achieved a moderately lower recall rate of 79.67\%, indicating the proportion of correctly predicted edges to the total actual edges. Our manual graph analysis showed that the lower recall is attributed to AndroByte’s D2CFG focus on privacy-sensitive dataflows, thus overlooking the next methods (caller-called edges) that are not directly associated with sensitive data. Androguard's call graph, on the other hand, is more generic and thus includes all caller-callee edges irrespective of the dataflow. Using a standard F1-score that balances precision and recall, AndroByte achieved an accuracy of 85.25\%. But given that the actual number of edges based on Androguard's ground truth may likely be more generic and those edges that are not part of the data-flow-aware call graph may be missed, we believe a F$\beta$-score for the accuracy metric will be more appropriate. In general statistics, the F$\beta$-score with $\beta<1$ is used when precision is more critical than recall. In our case, the false positives (Edges that appear in \( G_{\text{AB}} \) but are not found in \( G_{\text{AG}} \)) potentially signify dataflow edges which we should not be missed, while the false negatives (Edges that are missed by \( G_{\text{AB}} \) but are found in \( G_{\text{AG}} \)) are potentially edges that are not necessarily related to the dataflow. Thus, if we weigh the precision twice as much as recall and set $\beta$ = 0.5, the weighted harmonic mean will be 89.04\%.  

\begin{mdframed}[backgroundcolor=gray!10]
In summary, AndroByte can effectively generate accurate dataflow callgraph from bytecode summarization using LLM's reasoning and prompt engineering.
\end{mdframed}

\subsubsection{RQ3- How explainable is AndroByte’s method analysis in tracking sensitive dataflow and revealing potential data leaks in Android applications?} \label{tab:rq3} 
In this evaluation, we examine AndroByte's ability to contextualize bytecode instructions in natural language and evaluate the explainability of its summarizations in identifying sensitive dataflows. We utilize G-Eval with GPT4, a recently introduced evaluation framework by Liu et al. \cite{liu2023g}, which is specifically designed for open-ended natural language generation. This metric captures both semantic alignment and factual accuracy, and has demonstrated superior correlation with human judgments when compared to other LLM-based evaluation techniques. G-Eval is increasingly recognized \cite{chen2024driving,shayegani2023survey,kamoi2024can,yu2024llm} as a reliable metric for assessing the quality of generated content.

\begin{table}[h]
\centering
\caption{G-Eval Metrics for Generated Summaries}
\label{tab:geval-metrics}
\footnotesize
\begin{tblr}{
  width = \linewidth,
  colspec = {Q[1.3]Q[1]Q[1]},
  column{2} = {c},
  column{3} = {c},
  hline{1,8,9} = {-}{0.08em},
  hline{2} = {-}{},
}
\textbf{G-Eval Metric} & {\textbf{DroidBench}\\\textbf{Mean ± SD}} & {\textbf{UBCBench}\\\textbf{Mean ± SD}} \\
Coherence              & 4.65 ± 0.46                                & 4.66 ± 0.43                               \\
Consistency            & 4.10 ± 0.92                                & 4.30 ± 0.90                               \\
Relevance              & 3.45 ± 0.97                                & 3.40 ± 0.87                               \\
Fluency                & 4.65 ± 0.46                                & 4.66 ± 0.43                               \\
\hline
\textbf{Average per Dataset} & 4.21 ± 0.70                & 4.26 ± 0.66                      \\
\hline
\textbf{Overall Avg. Score} & \SetCell[c=2]{c}\textbf{4.24 ± 0.68} \\
\end{tblr}
\end{table}

In our evaluation, we define five core aspects of dataflow summarization quality: (1) identification of sensitive data types, (2) accuracy of data propagation, (3) correctness of sink function detection, (4) leakage inference capability, and (5) overall coherence and fluency. To quantify these aspects, we use the four evaluation dimensions defined in G-Eval. Specifically, consistency aggregates the first, second, and fourth aspects—capturing factual alignment and internal logic. Relevance reflects both data propagation accuracy and sink function detection, focusing on whether summaries emphasize key semantic content. Coherence and fluency correspond directly to our fifth aspect, assessing structural clarity and linguistic quality. This mapping enables us to compute G-Eval scores (on a 5-point scale) for AndroByte on both DroidBench and UBCBench, as reported in Table~\ref{tab:geval-metrics}, using the evaluation prompt provided in Appendix~\ref{appB}, Listing~\ref{lst:prompt1}.

\begin{figure*}[h]
\centering
\setlength{\fboxsep}{1pt}
\setlength{\fboxrule}{0.4pt}
\fcolorbox{black}{white}{%
  \includegraphics[width=\textwidth, height=0.5\textheight, keepaspectratio]{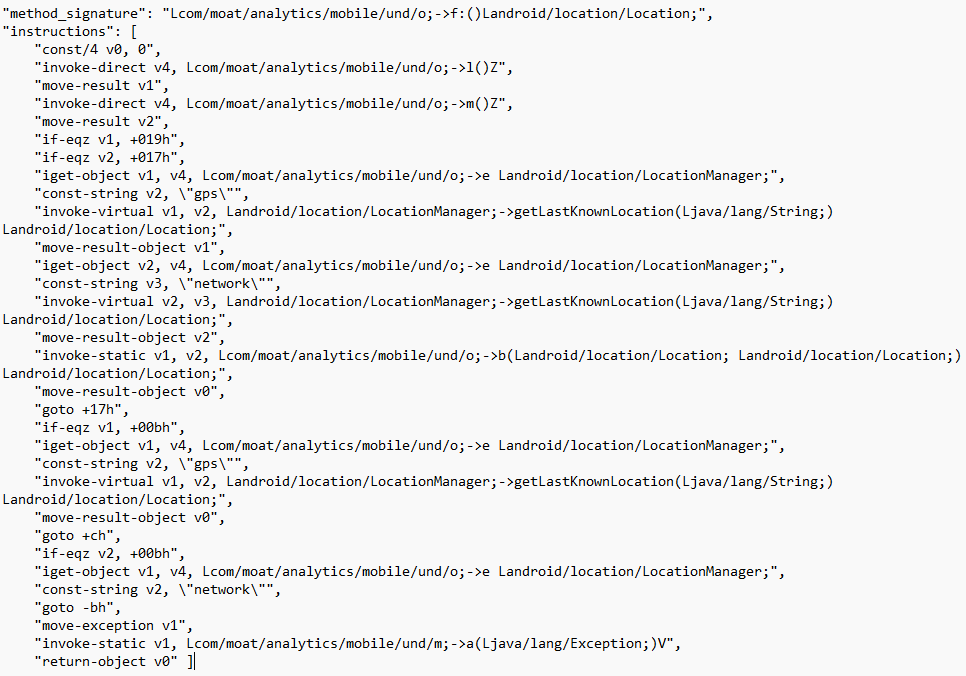}
}
\caption{Example – getLastKnownLocation() Call Site}
\label{fig:callSite}
\end{figure*}

To reduce model variance, all evaluations were repeated across 10 runs using a fixed temperature setting of 0.2 on the same dataset analyzed in RQ1. Since no publicly available ground truth method or graph-level summaries exist for Android apps, we derive ground-truth summaries by manually interpreting and distilling the code comments and inline documentation provided in each application of DroidBench and UBCBench benchmarks. While these summaries may not capture all implicit behaviors, they reflect a reasonable approximation of intended dataflow semantics and are used as reference baselines for evaluating AndroByte’s summarization performance. Table \ref{tab:geval-metrics} shows that AndroByte-generated summaries exhibit strong coherence and fluency ($4.65 \pm 0.46$ \&  $4.66 \pm 0.43$) across both datasets, reflecting clear and well-structured natural language outputs. Consistency also scores well ($4.10 \pm 0.92$ \& $4.30 \pm 0.90$), indicating reasonable factual alignment with the ground truth. However, the slightly lower relevance score ( $3.45 \pm 0.97$ \& $3.40 \pm 0.87$) may be partially attributed to the brevity and limited specificity of the ground truth summaries, which can hinder the evaluation model’s ability to fully capture the semantic coverage present in the generated outputs. AndroByte achieves an overall G-Eval score of 4.24, the average of its Coherence, Consistency, Relevance, and Fluency means.

\begin{table*}[t]
\centering
\caption{Call Site Summary}
 % \scriptsize
 % \footnotesize
\renewcommand{\arraystretch}{1.5} % Adjust row spacing
\begin{tabularx}{\textwidth}{|>{\raggedright\arraybackslash}X|>{\raggedright\arraybackslash}X|}
\hline
\textbf{Summary} & \textbf{Next Methods} \\ \hline
Sensitive user location data is retrieved from the \texttt{\textbf{LocationManager}} using \texttt{\textbf{getLastKnownLocation()}}methods for both GPS and network providers. The retrieved location data is then passed to the \texttt{b()} method and potentially forwarded to other methods. 
& 

1.\lstinline{ com/moat/analytics/mobile/und/o;->
b(android/location/Location;android/location/Location;)
android/location/Location;}
2.\lstinline{com/moat/analytics/mobile/und/m;->
a(Ljava/lang/Exception;)V}
 \\ \hline
\end{tabularx}
\label{tab:nextMeth}
\end{table*} 

\begin{table*}[h]
\centering
%\scriptsize
% \footnotesize
\caption{Next Method Summary}
\renewcommand{\arraystretch}{1.5} % Adjust row spacing
\begin{tabularx}{\textwidth}{|>{\raggedright\arraybackslash}p{0.45\textwidth}|>{\raggedright\arraybackslash}X|}
\hline
\textbf{Method Signature} & \textbf{Summary} \\ \hline
\lstinline{com/moat/analytics/mobile/und/o;->
b(android/location/Location;android/location/Location;)
android/location/Location;}
& 
"Method does not originate, store, or pass sensitive user personal data. No sink detected."
 \\ \hline
\lstinline{com/moat/analytics/mobile/und/m;->a(Ljava/lang/Exception;)V}
& 
"Sensitive API call detected: invoke-static \lstinline{com/moat/analytics/mobile/und/w;->
a()com/moat/analytics/mobile/und/w;}`.Data is logged using 
`invoke-static \lstinline{android/util/Log;->e(java/lang/String;java/lang/String;)I"}
 \\ \hline
\end{tabularx}
\label{tab:sink}
\end{table*}

\begin{table*}[h]
  \footnotesize
  \centering
  \caption{DataFlow Analysis Summary}
  \label{tab:summary}
  \renewcommand{\arraystretch}{1.5}
  \begin{tabularx}{\textwidth}{|>{\RaggedRight\arraybackslash}p{0.20\textwidth}|>{\RaggedRight\arraybackslash}X|}
    \hline
    \textbf{Parameters} & \textbf{Details} \\ \hline

    \textbf{Data Types} &
    Location \\ \hline

    \textbf{Overall Dataflow} &
    \begin{minipage}[t]{\linewidth}
      \vspace{2pt}%
      \begin{itemize}
        \item \textbf{Step:} Location retrieved from \lstinline!getLastKnownLocation()! 
        
              \textbf{Source Method:} \lstinline!Lcom/moat/analytics/mobile/inm/o;->f:()Landroid/location/Location;! 
        
              \textbf{Reasoning:} Sensitive data originated in this method.
        
              \textbf{Action:} Stored
        \item \textbf{Step:} Location passed to API method
        
              \textbf{Source Method:} \lstinline!Lcom/moat/analytics/mobile/inm/o;->b:(Landroid/location/Location; Landroid/location/Location;)Landroid/location/Location;! 
        
              \textbf{Reasoning:} Sensitive data is passed between methods.
        
              \textbf{Action:} Logged
      \end{itemize}
      \vspace{2pt}%
    \end{minipage}
    \\ \hline

    \textbf{All Sinks} &
    \lstinline!com/moat/analytics/mobile/inm/o;->a:(Landroid/location/Location;)V! 
    \\ \hline

    \textbf{Complete Dataflow} &
    \begin{minipage}[t]{\linewidth}
      % \vspace{2pt}%
      \lstinline!com/moat/analytics/mobile/inm/o;->f:()android/location/Location;! $\rightarrow$\\
      \lstinline!com/moat/analytics/mobile/inm/o;->b:(android/location/Location; android/location/Location;)android/location/Location;! $\rightarrow$
      \lstinline!com/moat/analytics/mobile/inm/o;->a:(android/location/Location;)V! \par
      \textbf{Reasoning:} Location data is retrieved, passed to API method, and then logged.
      \vspace{2pt}%
    \end{minipage}
    \\ \hline

    \textbf{Label} & Leak \\ \hline
  \end{tabularx}
\end{table*}

Furthermore, we manually inspected the output of AndroByte on some real-world apps from the 300 real-world apps in RQ2 (as no ground truth is available for systematic evaluation) and observed that it generates detailed summaries, identifies dataflows via dynamic call graphs, and detects sinks without relying on predefined rules and/or known sink lists. Figure \ref{fig:callSite} presents an example from the "myfitnesspal" app. Given the sensitive API getLastKnownLocation(), AndroByte identifies the call site as \lstinline{Lcom/moat/analytics/mobile/und/o}. Tracing the dataflow determines that the variables \texttt{v1} and \texttt{v2} hold the retrieved location data. This location data is subsequently passed to the \texttt{b()} function, which then further processes the data in the \texttt{a()} function. The \texttt{a()} function logs the data or handles exceptions.

Table \ref{tab:nextMeth} provides AndroByte's summary of the callsite (root node) and the list of next methods, as determined through summarization, which was dynamically added to the graph with edges connected to the root node. Table \ref{tab:sink} demonstrates how these methods are further summarized. Notably, the summary reveals that the \texttt{b()} function does not propagate the sensitive data. In contrast, the \texttt{a()} function does, passing the data to the Android logging utility, which signifies a potential data leak into the Android Log. Table \ref{tab:summary} provides a detailed output of a single analysis, starting at the call site. It includes summarization, recursive determination of the next methods, further summarization to identify potential sinks, or the determination of the next method. This process continues iteratively. This output, as illustrated, provides an overview of the dataflow, including reasoning, the identified sink, the complete dataflow, and a label indicating whether sensitive data is leaked. This evaluation validates the comprehensiveness of the summaries through logical reasoning and explainability, while highlighting AndroByte's ability to deliver a detailed, interpretable description of the data leak propagation and sink. A detailed heat map analysis of sensitive data handling by category is presented in Appendix \ref{appC} as Figure \ref{fig:heatmap}, highlighting the prominent exposure of location data compared to other data types.

\begin{mdframed}[backgroundcolor=gray!10]
In summary,  AndroByte demonstrates practical explainability in tracking sensitive data propagation and revealing potential leaks in mobile applications.
\end{mdframed} 
\vspace{0.2cm}

\section{Discussion}\label{discussion}%\vspace{-0.3cm}
 % \vspace{-0.3cm}
The evaluation of AndroByte underscores its effectiveness in dynamically generating automated D2CFG using language model reasoning and prompt engineering. A defining feature of AndroByte is its implementation of next-method tracking, which enhances its ability to detect dataflows from sensitive sources while concurrently building caller-callee relationships. It also validates the robustness of our bytecode summarization technique for method analysis, demonstrating the comprehensiveness, practical explainability, and accuracy of the generated summaries. This targeted approach enables precise analysis of user data collection processes, effectively identifying potential sensitive data flows leading to user data leaks in Android applications with accuracy as high as existing techniques but with much better adaptability. Utilizing bytecode summarization, AndroByte expands its contextual understanding and eliminates the need for predefined sinks, addressing the limitations and algorithmic complexities of traditional static taint analysis systems.

\textbf{Hallucination Mitigation:}
Hallucination concerns are addressed and integrated inherently into the design of AndroByte with practical mitigation strategies. First, the analysis process begins strictly from source methods of a decompiled APK that invoke sensitive APIs thereby avoiding the introduction of non-existent entry paths. Second, during method analysis, occasional hallucinations may occur, such as the generation of non-existent methods or invalid signatures by the model in the returned NEXT METHODS list. To address this, AndroByte performs validation against a statically extracted ALL METHODS list generated during the app decompilation. Any hallucinated method not present in this list is discarded and excluded from graph construction or further recursive method analysis. This lookup acts as a sanity check, ensuring only valid methods propagate in the graph. Third, once the full dataflow graph is generated, it is cross-validated against a call graph generated by AndroGuard described in Section ~\ref{tab:rq2}. For identifying sink detection and leak inference, AndroByte enforces the invariant that all leaks must appear as leaf nodes in the graph. In the explainability component, which summarizes the analysis, AndroByte uses (1) a consistency metric composed of: (i) accuracy of sensitive data type identification (ii) correctness of data propagation, and (iii) leak inference accuracy; and (2) a relevancy metric, which is composed of data propagation accuracy and sink function detection, as described Section ~\ref{tab:rq3}.

%\subsection{Limitations \& Future Work}
\textbf{Limitations \& Future Work:} Although AndroByte performs effectively on real-world data, DroidBench and UBCBench samples, it is not without its drawbacks, as itemized below: (1) Dependency on Bytecode Quality - While AndroByte effectively analyzes bytecode, heavily obfuscated or minified code can hinder its ability to generate accurate summaries or trace dataflows. Furthermore, apps that load code dynamically at runtime may bypass static analysis, leading to incomplete dataflow graphs. Although this is an important limitation, it is generic to all static analysis techniques irrespective of the algorithm applied. Given the huge overhead of dynamic analysis techniques, especially in taint analysis, tools like AndroByte are still very well needed in the community and cannot be completely replaced. (2) Dependency on LLM for Reasoning - AndroByte relies heavily on the reasoning capabilities of LLMs for bytecode summarization and graph generation. The resulting dataflow graph and analyses could be inaccurate if the LLM generates incorrect or irrelevant summaries due to context misinterpretation or insufficient domain knowledge, and without a robust validation mechanism, such errors may propagate and undermine reliability in critical scenarios. Although AndroByte performed well on three datasets, the robustness of AndroByte should be validated with much larger datasets. Additionally, introducing human reviewers for critical analyses is essential in real-world deployment. This could involve sampling LLM-generated summaries for manual validation. (3) Performance Overhead - The integration of language models for bytecode summarization is computationally expensive, particularly when analyzing large or complex applications. Real-world scenarios with a large volume of apps or highly interconnected call graphs may result in increased processing time and memory usage.
Furthermore, AndroByte's analysis duration significantly depends on the selected LLM’s context window size. In our experiments, we utilized a context window of 40,000 tokens with the local model gemma3:l, which notably improved AndroByte’s overall performance and scalability. The performance evaluation of AndroByte including method analysis, graph generation, sink detection, and summary generation shows that for real-world applications ranging in size from 2MB to 670MB (average 45MB), the average analysis time was approximately 2.5 minutes per application. While this runtime is acceptable for static analysis, further improvements are possible in real deployment scenarios with more advanced hardware, enabling even faster processing. For future work, we aim to conduct a human validation study to assess the accuracy and efficacy of the generated summaries in practical scenarios. We also intend to strengthen reliability by incorporating advanced LLM output validation mechanisms, such as confidence estimation, and by testing with larger datasets and diverse models, including GPT-NeoX, OPT, as well as frontier models, to explore performance variations and scalability. Furthermore, we aim to enhance AndroByte by incorporating static and dynamic analysis techniques, enabling a more comprehensive view of runtime data flows and improving the detection of privacy leaks.

%\vspace{-.3cm}
% \input{ACSAC 2025/Sections/2.relatedwork}
\section{Conclusion}\label{conclusion}
% \vspace{-0.3cm}
This study presents AndroByte, an innovative approach to privacy analysis that combines LLM-driven reasoning and bytecode summarization for method and data flow analysis. AndroBytes's design eliminates the reliance on predefined sinks and propagation rules, enabling automated detection of data leaks directly from the bytecode. Its dynamic dataflow call graph generation (D2CFG) component constructs a recursive and iterative representation of the program's behavior, connecting methods through caller-callee relationships and tracing sensitive data to identify potential sinks. The evaluation of AndroByte on real-world applications and benchmark datasets demonstrates its effectiveness in graph generation and identifying sensitive data propagation with high accuracy. Additionally, our evaluation highlights the comprehensive summary and practical explainability of the privacy analysis. By integrating AI reasoning, bytecode summarization, and recursive graph construction, AndroByte provides an effective, explainable, and adaptable technique for privacy analysis in Android applications, setting a new standard for automated privacy analysis tools.\\

\balance
\bibliographystyle{IEEEtran}
\bibliography{ref}

\newpage
\onecolumn
\begin{appendices}
% \section{Prompt Examples}\label{appA}  
% \vspace{-0.8cm} 

\noindent\begin{minipage}{\textwidth}
\section{Prompt Examples}\label{appA}
\vspace{-0.3cm}
\end{minipage}

\begin{figure}[h]
\begin{lstlisting}[frame=tlrb, caption={Example- getLastKnownLocation() Call Site}, label={lst:prompt}, basicstyle=\footnotesize\ttfamily, tabsize=2]      
   prompt = (
    "You are an expert in analyzing Android bytecode instructions. Your task is to trace how sensitive user data is originated, "
    "moved through registers, passed between methods, and possibly reaches sinks (e.g., logging, network, or storage).\n\n"
    "**Chain of Thought Process:**\n\n"
    "**1. Understand Context:**\n"
    f"- Previous Summary: {previous_summary}\n"
    f"- Method Signature: {method.get('method_signature', '')}\n"
    f"- Bytecode Instructions: {instructions_text}\n"
    "- Goal: Output JSON with 'Summary' and 'Next Methods'.\n\n"
    "**2. Identify Data Origin:**\n"
    "- Look for sensitive API calls (e.g., location, contacts, device ID).\n"
    "- Note data type, origin method, and the register it's stored in.\n"
    "- If no origin, check if sensitive data may come via parameters (from `Previous Summary`).\n\n"
    "**3. Track Data Storage:**\n"
    "- If sensitive data found, trace its flow (via `move-*`, `iput-*`, `sput-*`, etc.).\n\n"
    "**4. List Invoked Methods:**\n"
    "- Extract full method signatures from invoke-* calls.\n"
    "- Note which are passed sensitive registers.\n\n"
    "**5. Filter Next Methods:**\n"
    "- Exclude: `Landroid/*`, `Landroidx/*`, `Lkotlin/*`.\n"
    "- Only keep directly invoked methods.\n"
    "- If none left, use `[]`.\n\n"
    "**6. Detect Sinks:**\n"
    "- Check if sensitive data is passed to sinks like:\n"
    "  - Logging \n"
    "  - Network Transmission \n"
    "  - Storage \n"
    "- Return statements are not sinks.\n\n"
    "**7. Finalize 'Next Methods':**\n"
    "- If sink is hit with sensitive data, set `Next Methods` to `[]`.\n"
    "- Otherwise, keep filtered method list.\n\n"
    "**8. Construct Summary:**\n"
    "- Describe origin, movement, and whether sensitive data was passed or leaked.\n"
    "- If none observed, state it clearly.\n\n"
    "### Output Format:\n"
    "```json\n"
    "{{\n"
    '    "Summary": "[Summary of analysis based on the thought process]",\n'
    '    "Next Methods": ["FullyQualifiedClass->methodName:(params)returnType"]\n'
    "}}\n"
    "```\n\n"
    "- No markdown, code fences, or extra text.\n"
    "- Complete method signatures only.\n"
    "- JSON must be valid and standalone.\n\n"
    "**STRICT RULES:**\n"
    "1. Output only the JSON object. No explanation, markdown, or commentary.\n"
    "2. Method signatures: full, exact, no guessing, no truncation.\n"
    "3. `Next Methods = []` if a sink is hit.\n"
    "4. Do not reuse examples from the prompt.\n"
)

\end{lstlisting}
\end{figure}
\clearpage
\twocolumn 
\vspace{6cm}

\section{G-Eval Prompt Example}
\label{appB}  
\begin{figure}[h]
\begin{lstlisting}[
    frame=tlrb,
    caption={Prompt Used for Evaluation of Model-Generated Summaries},
    label={lst:prompt1},
    basicstyle=\footnotesize\ttfamily,
    tabsize=2
]
You are an AI assistant tasked with evaluating  model generated code analysis summary of an apk file privacy data leakage analysis. You will be provided with following context to analysis: 
- Ground Truth Summary: The result of manual expert analysis.
- Model Output: The output from a language model.

Your evaluation should focus on **privacy-related data flow** and **leakage behavior**. Do not penalize the model for providing more detail if it is **factually aligned**.

### Evaluation Dimensions (score each 1-5):
1. Data Type Identification - Does the model correctly identify sensitive data types (e.g., deviceId)?
2. Data Propagation Accuracy - Is the data movement (source -> transformation -> sink) accurately described?
3. Sink Function Match - Does the model correctly identify the final sink(s)?
   - Treat Java and bytecode references as equivalent.
4. Leakage Inference - Does the model correctly determine if data is leaked?
5. Coherence & Fluency - Is the output grammatically fluent and logically organized?

Please respond with valid JSON only using:
{
  "data_type_identification": <1-5>,
  "data_propagation_accuracy": <1-5>,
  "sink_function_match": <1-5>,
  "leakage_inference": <1-5>,
  "coherence_and_fluency": <1-5>
}
\end{lstlisting}
\end{figure}
\subsection*{G-Eval Metric Mapping}
The following mapping were used to convert structured evaluation scores, derived from privacy dataflow analysis in mobile applications.
\begin{table}[h]
\centering
\caption{G-Eval metrics for generated dataflow summary}
\begin{tabular}{p{0.25\linewidth} p{0.65\linewidth}}  % Adjust widths as needed
\toprule
\textbf{G-Eval Metric} & \textbf{Mapping Formula} \\
\midrule
Coherence   & Coherence \& Fluency (from model output) \\
Fluency     & Coherence \& Fluency (same as above) \\
Consistency & Mean of: Data Type Identification, Propagation Accuracy, Leakage Inference \\
Relevance   & Mean of: Propagation Accuracy, Sink Function Match \\
\bottomrule
\end{tabular}
\label{tab:geval-mapping}
\end{table}

 % \vspace{-1cm}

\section{Examples of Missed Data Leaks}
\label{appD}
\begin{figure}[h]
\begin{lstlisting}[
    frame=tlrb,
    caption={Example Where AndroByte Misses One of Two Sinks from the Same Source Method},
    label={lst:leak1},
    basicstyle=\footnotesize\ttfamily,
    tabsize=2,
    breaklines=true,
    breakatwhitespace=false,
    showstringspaces=false,
    literate={→}{{$\rightarrow$}}1
]
[
    {
        "Data Types Collected": [
            "Latitude",
            "Longitude"
        ],
        "Overall Data Flow": [
            {
                "Step": "Get location data",
                "Source Method": "Lde/ecspride/FactoryMethods1;->onCreate:(Landroid/os/Bundle;)V",
                "Reasoning": "Retrieves last known location using system service.",
                "Action": "Collected"
            },
            {
                "Step": "Log location data",
                "Source Method": "Lde/ecspride/FactoryMethods1;->onCreate:(Landroid/os/Bundle;)V",
                "Reasoning": "Logs latitude and longitude using Log.d.",
                "Action": "Logged"
            }
        ],
        "All Sinks": [
            "Landroid/util/Log;->d:(Ljava/lang/String;Ljava/lang/String;)I"
        ],
        "Complete Data Flow": [
            {
                "dataflow 1": "Lde/ecspride/FactoryMethods1;->onCreate:(Landroid/os/Bundle;)V → Landroid/location/LocationManager;->getLastKnownLocation:(Ljava/lang/String;)Landroid/location/Location; → Landroid/util/Log;->d:(Ljava/lang/String;Ljava/lang/String;)I",
                "Reasoning": "Sensitive location data is obtained and logged directly in the same method."
            }
        ],
        "Label": [
            "leak"
        ]
    }
]
\end{lstlisting}
\end{figure}

\clearpage

\section{Sensitive Data Actions: Summary-Level Insights}
\label{appC}  

 % \vspace{-0.3cm}
\begin{figure}[H]
\centering
\includegraphics[width=0.5\textwidth, height=6cm]{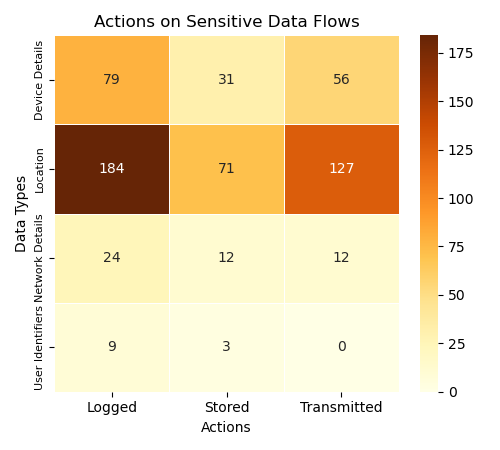}
\caption{Sensitive data Actions on data types}
\label{fig:heatmap}
\end{figure}

Based on our RQ3 results on real world apps, we categorize identified sinks into three types such as logging, storing, and transmitting. The heat map in Figure \ref{fig:heatmap} shows that the Location data emerges as the most frequently handled, with 184 instances of logging, 71 instances of storage, and 127 instances of transmission, highlighting its extensive use and potential vulnerability in Android applications. In contrast, actions on User Identifiers are minimal, with 9 instances of logging, 3 of storage, and no cases of transmission, indicating better adherence to privacy standards for this category. Device Details show significant logging and transmission, with storage being less frequent, while Network Details exhibit moderate handling across all actions. These findings highlight the importance of securing user location data, which has the highest percentage of sensitive data flow across both models.

\end{appendices}

\end{document}